\documentclass[fleqn,usenatbib]{mnras}

\usepackage{newtxtext,newtxmath}
\usepackage[T1]{fontenc}

\DeclareRobustCommand{\VAN}[3]{#2}
\let\VANthebibliography\thebibliography
\def\thebibliography{\DeclareRobustCommand{\VAN}[3]{##3}\VANthebibliography}

\usepackage{graphicx}	% Including figure files
\usepackage{amsmath}	% Advanced maths commands
\usepackage{xcolor}

\newcommand{\iras}{{IRAS 09149$-$6206}}

\newcommand{\ergs}{{\mbox{$\mathrm{erg\,s^{-1}}$}}}

\newcommand{\w}{UVW2}
\newcommand{\wb}{UVW1}
\newcommand{\m}{UVM2}

\newcommand{\orcid}[1]{\textsuperscript{\href{http://orcid.org/#1}{
\hskip2pt\includegraphics[width=8pt]{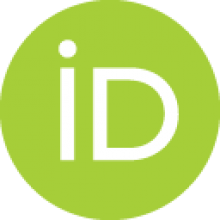}}}}

\newcommand{\py}{PyROA}

\title[Reverberation mapping of \iras]{On the nature of the continuum reverberation of X-ray/UV and optical emission of \iras.}

\author[Gonzalez-Buitrago et al.]{D. H. Gonz\'{a}lez-Buitrago\orcid{0000-0002-9280-1184},$^{1}$\thanks{E-mail: dgonzalez@astro.unam.mx}
Ma. T. Garc\'ia-D\'iaz\orcid{0000-0002-9772-5555},$^{1}$
F. Pozo Nu{\~n}ez\orcid{0000-0002-6716-4179},$^{2}$
and Hengxiao Guo\orcid{0000-0001-8416-7059}$^{3}$
\\
$^{1}$Universidad Nacional Aut\'onoma de M\'exico, Instituto de Astronom\'ia, AP 106,  Ensenada 22860, BC, M\'exico\\
$^{2}$Astroinformatics, Heidelberg Institute for Theoretical Studies, Schloss-Wolfsbrunnenweg 35, 69118 Heidelberg, Germany\\
$^{3}$Key Laboratory for Research in Galaxies and Cosmology, Shanghai Astronomical Observatory, Chinese Academy of Sciences, 80 Nandan Road,\\ Shanghai 200030, People’s Republic of China
}

\date{Accepted XXX. Received YYY; in original form ZZZ}

\pubyear{2023}

\begin{document}
\label{firstpage}
\pagerange{\pageref{firstpage}--\pageref{lastpage}}
\maketitle

% Abstract of the paper
\begin{abstract}
We present the results of a continuum reverberation mapping study of the radio-quiet Seyfert 1 galaxy \iras.
The analysis was performed using X-ray, UV and optical observations made with the {\em Swift} telescope between January and December 2017.
The time delays between different light curves were measured using three different algorithms: PyI$^2$CCF, PyROA and JAVELIN.
Our results show that the time delays increase with wavelength after $\tau\propto \lambda^{4/3}$, as predicted for a geometrically thin and optically thick accretion disc, but only after accounting for significant diffuse continuum emission from the broad line region.
However, the measured size of the accretion disc can be up to five times larger than that predicted by standard theory.
To our surprise, the strong increase in soft X-ray fluxes is delayed by about 15 days compared to the optical UV fluctuations, which challenges the prediction of the lamp-post model. Our analysis of the X-ray variability reveals the presence of a non-variable spectral component at 0.3-6.0 keV along with variable excess emission at 2.0-3.0 keV, which could be partly related to relativistic reflection in the inner region of the accretion disc.
\iras\ joins the list of objects for which the traditional lamp-post model cannot explain the observed time delays. A scenario that incorporates other geometric considerations into the lamp-post model, e.g. an extended corona along a scattering source, might be better suited to explain the observed long time delays.

\end{abstract}

% Select between one and six entries from the list of approved keywords.
% Don't make up new ones.
\begin{keywords}
accretion, accretion discs –galaxies: active –galaxies: Seyfert –quasars: individual: \iras
\end{keywords}

%%%%%%%%%%%%%%%%%%%%%%%%%%%%%%%%%%%%%%%%%%%%%%%%%%

%%%%%%%%%%%%%%%%% BODY OF PAPER %%%%%%%%%%%%%%%%%%

\section{Introduction}

Active galactic nuclei (AGN) are one of the most energetic and luminous phenomena in the universe, with luminosities that can reach up to $\sim10^{47}$ erg s$^{-1}$ with UV emission as its peak. 
This emission in the entire electromagnetic spectrum is believed to be caused by the release of energy from material falling spirally into a super massive black hole (SMBH) with masses up to $10^9$ M$\sun$. 
The physics of this process is usually discussed in the context of the standard theory of a geometrically thin and optically thick accretion disc (\citealt{1969Natur.223..690L}; \citealt{1972A&A....21....1P}; \citealt{SS73}; \citealt{1973blho.conf..343N}; hereafter referred as the standard AGN accretion disc theory).

AGN have a wide range of amplitude variability that can be monitored at all wavelengths. The rapid variability indicates that the variability time is shorter than the time it takes for light to cross the AGN, implying that their inner regions are very compact and cannot be resolved with current telescopes. 
Only a few exceptions for nearby objects exist, including M81 \citep{EHT19}, 3C 273 and IRAS 09149-6206 \citep{Gravity:2018, GRAVITY20}.
Therefore, additional tools which are independent on spatial resolution are needed to allow the study of the inner regions and identify the processes responsible for the observed variability. 
One of the best tools to date for accomplishing this task is reverberation mapping (RM, \citealt{Blandford82}).
The RM technique uses spectroscopy (\citealt{gask86}; \citealt{pet93}; \citealt{kaspi2000}; \citealt{bentz13}) or photometry (\citealt{cherepa73}; \citealt{haas11}; \citealt{pozo2012}; \citealt{chelo12}) to measure the time delay $\tau$ between the triggering continuum variations of the accretion disc and the response of regions located further away (e.g. the broad line region and dust torus). 
The delay gives to a first approximation the size of the reprocessed region, $R\sim \tau c$, where $c$ is the speed of light.
RM can also be applied to study the accretion disc in details. 
For example, in the standard theory of AGN accretion discs, the disc is optically thick and geometrically thin with a temperature profile given by $T(R)\propto R^{-3/4}$. 
Given that the time delay is $\tau \sim R/c$ and according to Wien's law, $\lambda \propto 1/T$, leads to the relation between the time delay and the wavelength of the form $\tau \propto \lambda^{4/3}$. 
This relation allows to estimate the radial extent of the accretion disc (see \citealt{2021iSci...24j2557C} for a review).

\begin{table*}
	\centering
	\caption{Log of time-resolved {\em Swift} observations and \textbf{light curve statistics for each band.}}
	\label{tab:logObs}
	\begin{tabular}{lccccccccc} % four columns, alignment for each
		\hline\hline
		Date &    Instrument & Band& Exposure Time & \# points & Time Span & Average & Median & Excess & $R_{\rm max}$\\
		     &         &  &  [s]  &  & [days] & Obs. & Obs. & variance & Flux Ratio\\
       (1) & (2) & (3) & (4) & (5) & (6) & (7) & (8) & (9) & (10) \\
		\hline
    	2017/01/06 - 2017/12/26 &  XRT  & 0.3 - 10 keV & 1500 - 2500   \\
		2017/01/06 - 2017/12/26 & UVOT  & \w & 196 & 86 & 354.49 & 4.17 & 3.50 & 0.02 & 1.32 \\
		 &   & \wb & 117 & 93 & 354.49 & 3.85  & 3.31 & 0.01 & 1.24 \\
		 &   & \m & 290  & 136 & 354.49 & 2.62 & 2.45 & 0.01 & 1.49 \\		
		 &   & U & 9.6 & 185 & 354.49 & 1.92 & 0.10 & 0.02 & 1.38 \\		
		 &   & B & 9.6 & 183 & 354.49 & 1.94 & 0.10 & 0.01 & 1.34 \\		
		 &   & V & 9.6 & 162 & 354.49 & 2.20 & 0.53 & 0.01 & 1.40 \\		
   \hline
		Binning 2 &   & &  &  &  &  &  &  &  \\
		 &   & U & 38.4 & 30 & 334.14 & 11.52 & 10.73 & 0.0074 & 1.18 \\		
		 &   & B & 38.4 & 30 & 338.74 & 11.68 & 10.69 & 0.0040 & 1.15 \\		
		 &   & V & 38.4 & 27 & 339.41 & 13.05 & 11.71 & 0.0001 & 1.12 \\		
   \hline
  \end{tabular}
  \noindent
{\em Notes:}
(5) Number of data points in the light curve. (6) The time span in days. 
(7)--(8) Average and median interval between observations. (9) excess variance
(10) ratio of maximum to minimum flux.
\end{table*}

%\w, \wb, \m, U, B, V & $\sim$196/117/290/9.6/9.6/9.6

There is a strong temporal relationship between the variability observed in X-rays, UV/optical and infrared \citep{Edelson96,Edelson17,Edelson19}. 
This connection is assumed to be caused primarily by a central, highly energetic X-ray-emitting source known as the corona. 
The corona is located at a certain height $h$ of the SMBH and illuminates the accretion disk. 
The disc absorbs, reprocesses, and remits this emission at different wavelengths, but with a time delay that corresponds to the travel time of light between the corona and the local region of the accretion disk that re-emits the emission. 
Therefore, the observed variability in wavelengths such as UV/optical are mainly driven by flux variations emitted in X-rays from the corona, this process is known as lamp post model (\citealt{Krolik1991}; \citealt{2005ApJ...622..129S};  \citealt{2007MNRAS.380..669C, 2021iSci...24j2557C}).

AGN often exhibit an excess of soft X-ray emission, typically peaking at 1-3 keV, over the power-law continuum that dominates the hard X-ray emission. One of the main explanations for this excess is the presence of a warm absorber, i.e. a partially ionised gas that absorbs the harder X-rays but transmits the softer ones. The warm absorber can be modelled as a series of absorbing layers with different ionisation states and column densities (e.g. \citealt{2007ApJ...659.1022K}). Alternatively, Comptonisation of soft photons by a hot corona above the accretion disc may contribute to the excess (e.g. \citealt{1991ApJ...380L..51H}). Another possible cause is the reflection of the hard X-ray continuum from a cold disc surface, which produces a fluorescent iron K$\alpha$ line and a Compton hump, both of which can enhance the soft X-ray fluxes (e.g. \citealt{1989MNRAS.238..729F}).

However, the exact origin of the variability and the reflective source are still under debate.
Currently, RM studies of multiple AGN have increased considerably, finding results that are helping to resolve the paradigms that govern these systems \citep{Edelson15,Fausnaugh16,McHardy18,pozo19,Cackett2020,hernandez2020}. 
Among their main results are: a) the time delay obtained for different wavelengths from X-rays to infrared follow the relationship $\tau \propto \lambda^{4/3}$ as expected for the standard AGN accretion disc theory, b) estimates of the size of the accretion disc suggest that they are a factor of 2-5 or even 11 times larger than predicted by the standard AGN accretion disc theory \citep{2023arXiv230207342K}, although solutions to this problem which involves underestimated black hole masses (\citealt{pozo19}) and internal AGN extinction (\citealt{gask23}) has been proposed, c) the delays found for the light curves in the $U/u/r$ bands (3467\AA, 3540\AA, 6215\AA) depart from the trend shown by the delays of the other bands that follow $\tau \propto \lambda^{4/3}$. 
This is likely caused by contamination from diffuse continuum emission of the BLR at these wavelengths \citep{Korista2001,Korista19, Lawther18, Guo2022} or perhaps to other under-appreciated non-disk components (\citealt{che19}), and d) the correlation of the variability between the X-ray and UV light curves is weaker than the correlation between the UV and optical light curves \citep{Edelson19}, thus raising doubts about whether X-ray emission is the driver of the variability in longer wavelengths.

In this paper, we present the results of a RM study on the AGN \iras\, performed using X-ray, UV and optical observations taken with the \textit{Neil Gehrels Swift Observatory} (hereafter referred to as {\em Swift} telescope).
At a redshift of 0.057 \citep{Perez89}, \iras\, is classified as a radio-quiet Seyfert 1 galaxy \citep{Perez89,Cram92}, however the exact morphology of the host galaxy is not yet known \citep{Kishimoto2011, Burtscher2013,LopezGonzaga2016}. 
\iras\, has been one of the AGN for which the $Br\gamma$ broad emission line has been spatially resolved through interferometry \citep{Gravity:2018,GRAVITY20}. 
A BLR radius of 0.075 pc has been inferred similar to the results obtained by RM studies. 
It features a black hole mass of about $10^8$ M$\sun$, and the possible presence of a disc reflection \citep{Liebmann2018}. In the X-ray spectral analysis carried out by \cite{Walton2020}, it was found that a partially ionised warm absorbent was present in the low energy range. At high energies, they observed evidence of strong relativistic reflection, possibly originating from the inner part of the accretion disc.
The complex and intriguing variability patterns across multiple wavelengths make \iras\, an ideal target for continuum RM and provide the opportunity to study the accretion processes and the interaction between the disc and the BLR in an AGN with distinctive observational characteristics.

This article is structured as follows: Section 2 describes the observations and data reduction. Section 3, we describe the time series analysis.
In Section 4, we model the time delay spectrum, analyse flux variation gradient, and discuss the results within the context of the standard AGN accretion disc theory.
Conclusions are presented in Section 5. 

\section{Observations and data reduction}
\label{sec:obs}
The data was gathered from the {\em Swift} Telescope database, which conducted continuous multi-wavelength X-ray, UV, and optical observations of \iras. The observations were made between January and December 2017, with a total of 88 visits and an exposure time of 1 $ks$ per visit. 
Simultaneous observations were carried out with both the UV and optical telescope \citep[UVOT,][]{Roming2005} and its 6 filters (\w, \wb, \m, $U$, $B$, $V$)  with to 3 days cadence observations and the {\em Swift} X-rays \citep[XRT,][]{Burrows2005}, using window timing mode. 
A summary of the observations and the light curve statistics for each band are given in Table~\ref{tab:logObs}.

For the reduction of the UVOT telescope data and the extraction of the light curves in its 6 filters we used the standard procedure via the ftools package \citep{Blackburn1995}.
During our observations, the exposure time of the V-band was not sufficient to obtain a higher signal-to-noise ratio (SNR) for individual measurements. Since 2019, a new mode (0x224c) has been used in the Swift-based continuum RM programs, which increases the exposure time in the V and W2 bands. Since our observations preceded this change, we implemented a binning technique to mitigate noise effects and improve the reliability of the light curves. Specifically, we used an averaging strategy in which two adjacent data points were combined into a single data point. This binning procedure effectively reduced the effects of individual measurement errors and fluctuations. In particular, we observed a remarkable improvement in SNR for the V-band from 54 to 76. The binned light curves have smoother profiles that provide a clearer and more coherent representation of the underlying patterns of variability. Furthermore, the values for the ratio of maximum to minimum flux ($R_{\rm{max}}$) and the excess variance (see Table~\ref{tab:logObs}) are very similar for the original light curve without binning and the binned light curve, with only a 20\% reduction in variability observed for the latter. This indicates that the binning procedure effectively attenuates the noise effects without significantly changing the amplitude of the variability in the $U$-, $B$- and $V$-band light curves. We discuss the effects of the light curve binning procedure on the determination of the time delay in Section \ref{sec:time_analysis}.

An aperture with a radius of 5 arc-seconds was used over the centre of the AGN.
At this stage, the flux or contamination of the contribution of the host galaxy was not removed since it is constant and does not affect the variability of the AGN light curves in any of the filters used.
However, disentangling the galaxy from the total flux is crucial to obtain an unbiased estimate of the AGN luminosity. 
A careful treatment of the subtraction of the host galaxy contribution is given in Sect. \ref{sec:FVG}. 
The extracted sky background was obtained from a ring around the source with internal and external radius of 10 and 20 arc-seconds, respectively.

X-ray light curves were extracted using the {\em Swift} XRT Products online tool\footnote{\url{https://www.swift.ac.uk/user\_objects/}} developed by the UK {\em Swift} Center.
More information can be found in \cite{Evans2007,Evans2009}. 
The tool produce light curves that are fully corrected by instrumental effects such as buildup, dead regions on the CCD, and vignetting. 
We separated our analysis in two sets of X-ray light curves divided into different energy ranges. 
In the first set, we consider five light curves in the soft X-ray (SX) between 0.3 and 3.0 keV and one light curve in the energy range of 3.0-10 keV called hard X-rays (HX). 
For the second group, we divided the 0.3-10 keV energy range into ten light curves (0.3-1.0, 1.0-2.0, 2.0-3.0, 3.0-4.0, 4.0-5.0, 5.0-6.0, 7.0-8.0, 9.0-10 keV). 
Each light curve was obtained using the snapshot binning method, which produces a measurement for each point on the spacecraft. 
The X-ray, UV and optical light curves are shown in Figure~\ref{fig:LCall}. 
Figure~\ref{fig:lc_xray} shows the HX light curves for the selected ten energy ranges.

\begin{figure*}
 	\includegraphics[trim=0.2cm 0.0cm 0.0cm 0.0cm, clip, width=17cm]{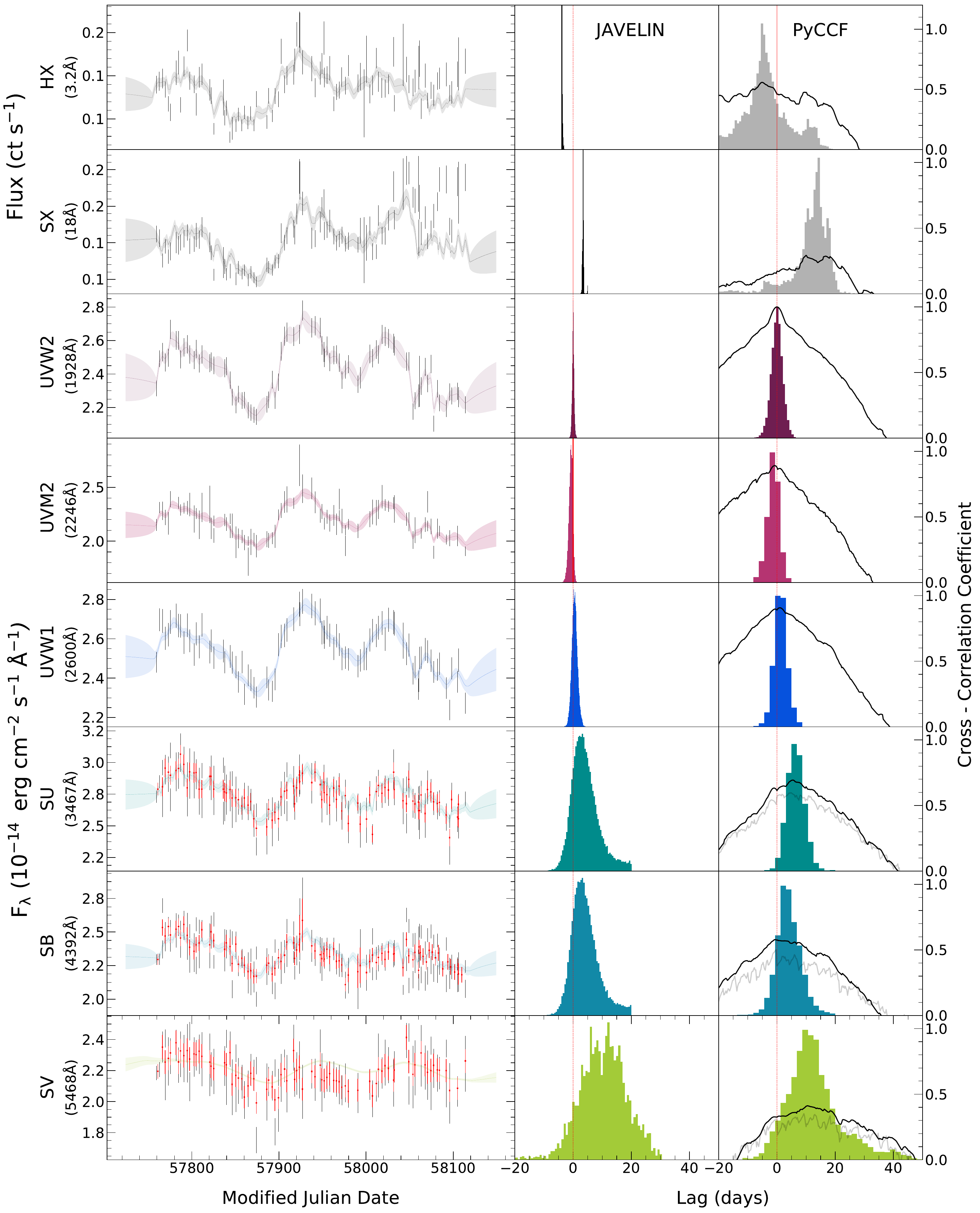}
\caption{X-ray/UV and optical light curves with time delay analysis. The two upper left panels show the X-ray light curves in energy ranges of 3.0-10 keV (HX-ray) and 0.3-3.0 keV (SX-ray). The lower left panels correspond to the light curves of the bands \w, \m, \wb, U, B, V. The original light curves (without binning) are shown in black, the binned light curves in red. The solid line in each left panel corresponds to the model made to the light curve with Javelin. The right panels show the histograms corresponding to the time delay obtained with JAVELIN (middle panel) and {\tt PyI$^2$CCF} (right panel). The dotted red line marks a zero time delay.}
   \label{fig:LagJavelin}
\end{figure*}

\begin{table*}
	\centering
	\caption{Time delay measurements with respect to the UVW2 band. Columns (1) correspond to the band name. Column (2) and (3) are the observed and the rest wavelength, respectively. Column (4) is the maximum correlation coefficient. Columns (5) and (6) are the PyI$^2$CCF delay peak and centroid, respectively. (7) Javelin delay and (8) PyROA delay.}
	\label{tab:lagTime}
	\begin{tabular}{lccccccc} % four columns, alignment for each
		\hline\hline
		Band & $\lambda_\mathrm{observed}$ & $\lambda_\mathrm{rest}$ & $r_\mathrm{max}$ & PyI$^2$CCF & PyI$^2$CCF & {\sc javelin} &  \py\\
		     &   &   &   & $\tau_{\rm peak}$& $\tau_{\rm cent}$ &$\tau_{\rm JAV}$&$\tau_{}$\\
		     &[\AA] & [\AA] &  & [days] & [days] & [days]    &  [days]   \\
		(1)     & (2) & (3) & (4) & (5) & (6) & (7) & (8) \\
		\hline
        \hline
		HX   & 3 & 2.88& 0.55 & $-4.56^{+8.08}_{-8.47}$ & $-5.02^{+9.90}_{-8.86}$ &$-4.59^{+5.14}_{-3.89}$   &  $\cdots$ \\[.15cm]
		SX   & 1.8 & 1.70& 0.33 & $17.03 ^{+7.15}_{-7.16}$ & $13.46^{+4.01}_{-4.44}$ &$5.13^{+5.37}_{-5.55}$  &  $\cdots$ \\[.15cm]
		X$_{5-10keV}$   & 1.8 & 1.70& 0.50 & $-6.34^{+8.69}_{-17.11}$ & $-6.58^{+7.78}_{-7.89}$ &$-5.94^{+9.65}_{-2.34}$   &  $\cdots$ \\[.15cm]
  		X$_{2.5-5keV}$   & 3.5 & 3.31& 0.51 & $-3.73^{+7.50}_{-7.15}$ & $-4.28^{+8.23}_{-7.75}$ &$-4.96^{+5.30}_{-4.52}$   &  $\cdots$ \\[.15cm]
  		X$_{1.25-2.5keV}$   & 7.8 & 7.37& 0.20 & $13.49^{+13.34}_{-4.90}$ & $13.10^{+4.99}_{-11.59}$ &$8.98^{+8.56}_{-9.55}$   &  $\cdots$ \\[.15cm]
  		X$_{0.3-1.25keV}$   & 23 & 21.75& 0.40 & $13.33^{+16.46}_{-4.38}$ & $13.25^{+4.76}_{-16.47}$ &$12.19^{+8.74}_{-14.85}$   &  $\cdots$ \\[.15cm]
%		SX   & 10 & $\cdots$ & $\cdots$ &$\cdots$ &$-2.8^{+0.3}_{-0.3}$   & & \\[.15cm]
		UVW2 & 1928 & 1824 & 1.00  &$0.00^{+0.80}_{-0.79}$ & $0.04^{+0.80}_{-0.79}$ & $0.02^{+0.94}_{-0.91}$ &  $0.00^{+0.00}_{-0.00}$ \\[.15cm]
  		UVM2 & 2246 & 2124 & 0.89 & $0.80^{+4.23}_{-4.43}$ & $-1.29^{+1.95}_{-2.00}$ & $-1.28^{+2.01}_{-2.01}$ &  $0.22^{+0.16}_{-0.31}$\\[.15cm]
        UVW1 & 2600 & 2459 & 0.90 & $1.25^{+2.16}_{-2.19}$ & $1.03^{+2.24}_{-2.09}$ & $0.41^{+0.53}_{-1.41}$ &  $0.96^{+0.51}_{-0.56}$ \\[.15cm]
        U & 3467 & 3047 & 0.70  & $6.36^{+3.30}_{-3.47}$ & $6.41^{+2.91}_{-2.97}$ & $5.65^{+2.60}_{-2.23}$ & $5.23^{+0.85}_{-0.86}$ \\[.15cm]
		B & 4392 & 3862 & 0.57  & $3.89^{+4.56}_{-4.97}$ & $3.72^{+3.44}_{-4.29}$ & $3.54^{+1.23}_{-1.12}$ & $3.40^{+1.47}_{-1.48}$
		\\[.15cm]
		V & 5468 & 4808 & 0.41 & $12.91^{+8.67}_{-11.67}$ & $11.60^{+7.36}_{-9.79}$ & $13.10^{+7.51}_{-7.55}$ &  $3.89^{+2.32}_{-2.34}$ \\[.15cm]	
		\hline
	\end{tabular}
\end{table*}

\section{Time series Analysis}
\label{sec:time_analysis}

The \w\, band (centred on 1928\AA\,) shows greater variability and spans a wavelength range representative of the continuum emitted by the accretion disc. Since it is also less affected by contamination from other emission components (such as the BLR), we chose this light curve as a reference to estimate the delay time between the light curves of the different bands (\w, \wb, \m, $U$, $B$, $V$). In addition, we also performed time-delay measurements for the binned $UBV$-band light curves (see Figure~\ref{fig:LagJavelin}). The time delay analysis was performed using three different algorithms: PyI$^2$CCF, PyROA and JAVELIN. Each algorithm uses specific configurations tailored to our analysis. PyI$^2$CCF uses a cross-correlation technique, PyROA implements a regularised optimal estimation approach and JAVELIN uses a damped random walk model to estimate the time delays.

Below we briefly describe each method and the specific configuration used in this work.

\subsection{PyI$^2$CCF}

The interpolated cross-correlation (ICCF; \citealt{1987ApJS...65....1G}) method is used to measure the time delay between two light curves by first interpolating them linearly and then shifting one curve by a specific time lag. 
The cross-correlation coefficient $r$ is then calculated at this lag and the lag with the highest correlation is found by searching over a range of values. 
Here we use the {\tt PyI$^2$CCF} code, which is a publicly available package\footnote{\url{https://github.com/legolason/PyIICCF}} \citep{Guo21}, to perform the ICCF lag analysis. 
This code implements the ICCF method and also includes a null hypothesis test to evaluate the reliability of the detected correlation. 
{\tt PyI$^2$CCF} offers other nonlinear interpolation methods but in this study, linear interpolation was used following the standard ICCF technique, which is identical to the {\tt PyCCF} method \citep{Peterson98,Sun18}. 
The lag search range in {\tt PyI$^2$CCF} was set to $\pm$60 days in the observed frame with a grid spacing of 0.5 days, which is sufficient for the 3-day cadence of the light curves. 
PyI${^2}$CCF employs the traditional flux randomization (FR) and random sub-sampling procedure to obtain the lags and uncertainties. 
This Monte Carlo method randomizes the flux measurements by their uncertainties (FR) and randomly chooses a subset of light-curve points (RSS) to build the centroid lag distribution. 
The median value and 1$\sigma$ range of this distribution serve as the centroid lag and error, respectively. 
We carried out 10000 FR/RSS iterations for each measurement. A single {\tt PyI$^2$CCF} realisation including the histograms from the FR/RSS distribution are shown in Figure~\ref{fig:LagJavelin}.

\begin{figure*}
 	\includegraphics[trim=0cm 0.0cm 0.1cm 0.1cm,clip,width=18cm]{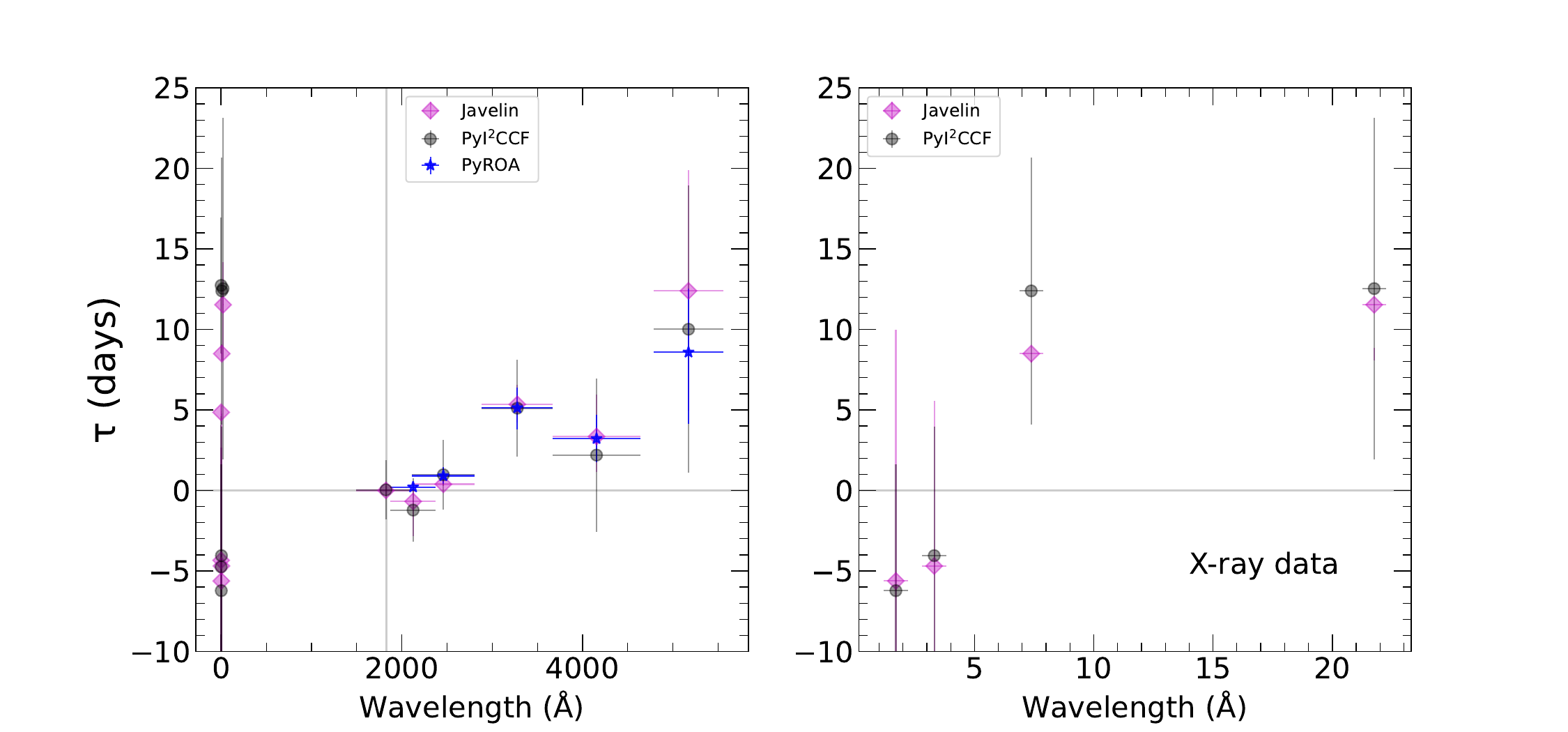}
\caption{Time delay spectrum. The left panel displays the X-ray to V-band time delays.
The panel on the right depicts the temporal delays between the soft and hard X-ray bands. The blue dots represent the time delays obtained by PyROA, the grey dots by {\tt PyI$^2$CCF} , and the pink dots by JAVELIN.
}
    \label{fig:lag_data}
\end{figure*}

\subsection{JAVELIN}

The JAVELIN code\footnote{\url{https://github.com/nye17/javelin}} (\citealt{2010ascl.soft10007Z}) assumes that the time-shifted signal is the result of the convolution between the driver and a top-hat transfer function, and estimates the errors using the Markov Monte Carlo (MCMC) approach. 
JAVELIN first interpolates the reference light curve assuming that the AGN variability is well described by a Damped Random Walk (DRW) model. 
Then it convolves the DRW model with the top-hat transfer function, shifts, and scales the resulting light curve to compare with the other bands. After performing this process several times through the MCMC procedure, it obtains the marginalised posterior distribution for the lag, the width of the top hat, the scale factor of the light curve, the time scale and amplitude of the DRW model. 
For our case, around 15000 MCMC sample-and-burn iterations were carried out to estimate the probability distributions of the delays and other parameters of the DRW model. 
Figure~\ref{fig:LagJavelin} shows the light curves together with the best fit obtained with JAVELIN, together with the histograms corresponding to the time delay distribution.

\subsection{PyROA}

The PyROA algorithm\footnote{\url{https://github.com/FergusDonnan/PyROA}} was developed to model the behaviour of AGN light curves \citep{Donnan2021} by a running optimal average ROA. PyROA models the light curves through an MCMC with a Bayesian approach that allows a convolution between the light curve obtained from the best fit and the transfer function, allowing for uniform, Gaussian and log-Gaussian blurring. This provides the flexibility to create well-defined models and even light curves with low observational sampling. We ran PyROA to fit the light curves of the \w, \m, \wb, $U$, $B$ and $V$ bands simultaneously with 100000 iterations. PyROA generates the reference light curve and calculates the parameters for the delay and the rms value of the delay distribution. The light curves together with the fits produced by PyROA including the histograms are shown in Figure \ref{fig:pyroa} in the appendix. 
A summary of the time delay measurements obtained with all methods can be found in Table~\ref{tab:lagTime}.

\section{Discussion}

\subsection{Light curve variability and time delays}
\label{lcandvariable}

The light curves of the six UVOT filters and X-rays in different energy ranges were extracted (see section~\ref{sec:obs}). 
They show a high variability with several peaks, as well as a correlation of variability between the different light curves (see Figure~\ref{fig:LCall}). 
The variability decreases with increasing wavelength, with greater variability in the \w-band, while little variability is observed in the V-band, which is also highly noisy. 
A correlation with the X-ray light curves is also observed, but it is weaker than that in the UV/optical bands. 
This behaviour is very similar to that reported for other AGN (e.g., \citealt{Fausnaugh16, Edelson17,Edelson19,hernandez2020,GonzalezB2022}).

We calculated the time delay for the X-ray, UV, and optical light curves using the algorithms JAVELIN, PyROA, and PyI$^2$CCF described in section~\ref{sec:time_analysis}. 
The \w- band was considered as a driver. Figure~\ref{fig:lag_data} shows the recovered time delay spectrum.
These measurements confirm that the UV and optical light curves are correlated and the delay increases with increasing wavelength.
This reflects the behaviour of the reprocessing model of an optically thick and geometrically thin accretion disc as described by the standard AGN accretion disc theory, with some exceptions: a) The presence of a clear excess is observed in the lag spectrum at the position of the U-band. This excess can be attributed to an additional contribution of diffuse continuum emission (DCE) from the BLR (\citealt{Korista19}). We discuss the DCE contribution in more detail in Section \ref{ADmodeling}, b) A negative time delay of about $-5\pm9.0$ days was obtained for the hard X-ray emission (3.0-10 keV), indicating that the driver light curve \w\ lags behind the hard X-ray emission. This finding aligns with the expectations of the reprocessing model, where the emission is expected to be delayed due to light travel time effects. In contrast, a positive time delay of 13$\pm4.0$ days was observed for the soft X-ray emission (0.3-3.0 keV), suggesting that the soft X-ray light curve lags behind of the driver. This observation contradicts the predictions of the lamp post model, which anticipates the soft X-ray emission to lead the UV/optical emission.

To better understand this behaviour, we extracted X-ray light curves in four energy ranges (0.3-1.25, 1.25-2.5, 2.5-5.0 and 5.0-10 keV) and quantified the delay with respect to the \w\, light curve.
In these measurements, negative delay were obtained for the 2.5-5.0 keV and 5.0-10 keV bands. 
For example, delays of -5.6$\pm$7.0 and -4.3$\pm$8.0 light days were found with the PyI$^2$CCF code. For the 0.3-1.25 and 1.25-2.5 keV bands, positive values of 13$\pm$5 and 13$\pm$4 were obtained with PyI$^2$CCF and very similar values with JAVELIN (see Figure~\ref{fig:lag_data} right panel). We discuss the implications of this long soft X-ray time delay in section \ref{ffpanalysis}. Table~\ref{tab:lagTime} shows a summary of the time delay results.

\subsection{Host subtracted AGN luminosity and mass accretion rate}
\label{sec:FVG}

To isolate the contribution of the AGN from the host galaxy, we applied the Flux Variation Gradient (FVG) method as described in \cite{pozo2012} for reverberation mapping data. 
In short, the total flux measured through different bands with the same photometric aperture comprise the variable flux from the AGN together with the constant flux from the host galaxy (including the contribution of the narrow-line region). 
The total fluxes follow a linear slope in a flux-flux diagram representing the colour of the AGN, while the colour of the host galaxy lies in a well-defined range (e.g. \citealt{2010ApJ...711..461S}).
The intersection between the slopes of the AGN and the host galaxy provides an estimate of the the host galaxy flux at the time of the observing campaign in the chosen photometric aperture. 

Before using the FVG method, the total fluxes must be corrected for Galactic foreground extinction. 
For this purpose, we used the recalibrated values obtained by \cite{2011ApJ...737..103S} from the dust extinction maps of \cite{1998ApJ...500..525S} and linearly interpolated (extrapolated) at the specific {\em Swift} wavelengths (Table \ref{fvgsumm}). For the extrapolation we have included the UV NED extinction value at 0.22um, as otherwise the UV fluxes (e.g. in UVW2) could be underestimated by about 50\%. For longer wavelengths (e.g. U,B,V) the effect is negligible (about 1\%).

\begin{figure}
  \centering
  \includegraphics[width=\columnwidth]{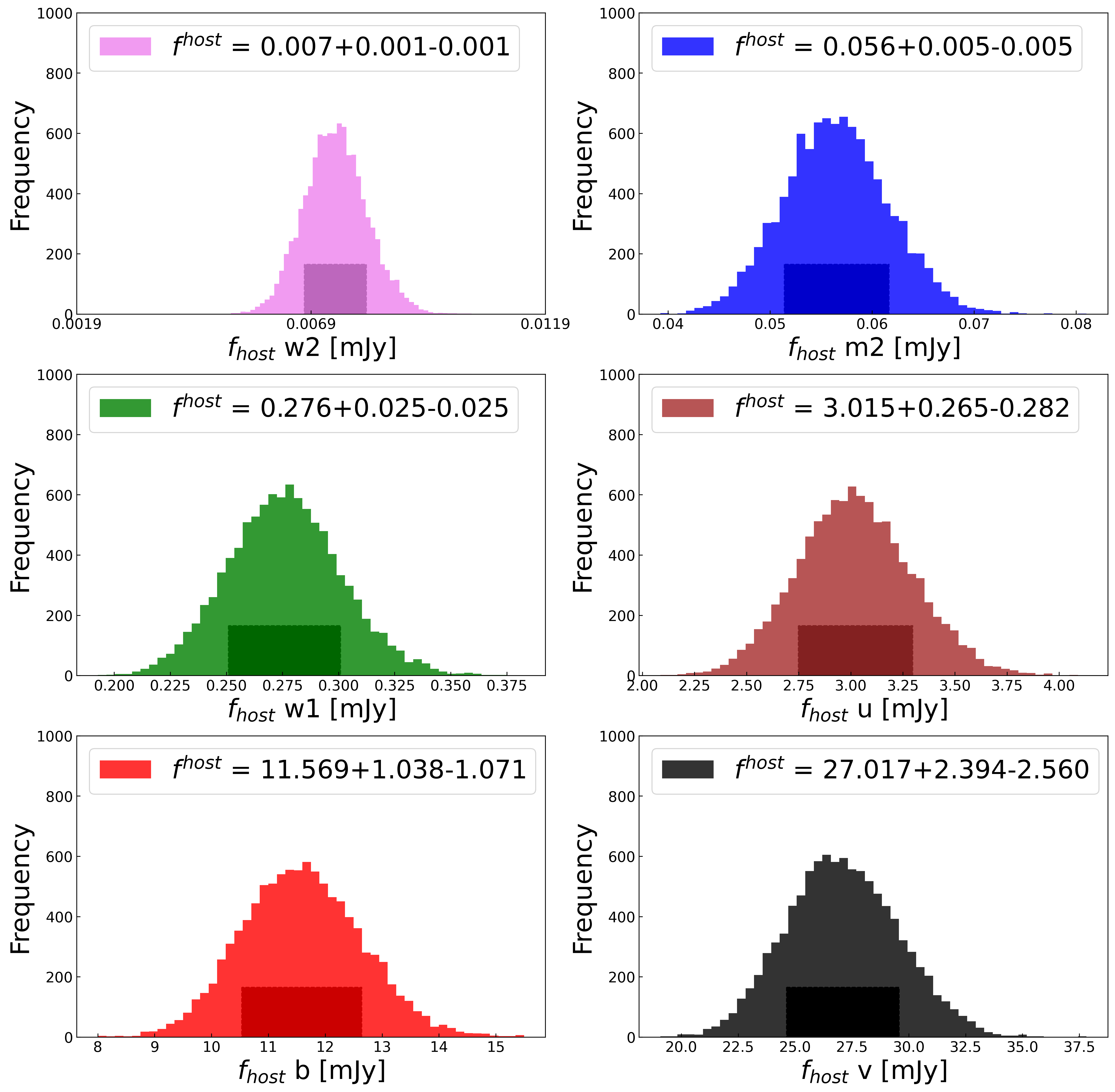}
\caption{PFVG distributions of host galaxy fluxes for bands \w, \m, \wb, U, B, and V. The shaded area marks the 68\% confidence range used to estimate the 1$\sigma$ uncertainty around the median of the distribution.}
\label{fvghost}
\end{figure}

\begin{figure}
  \centering
  \includegraphics[width=\columnwidth]{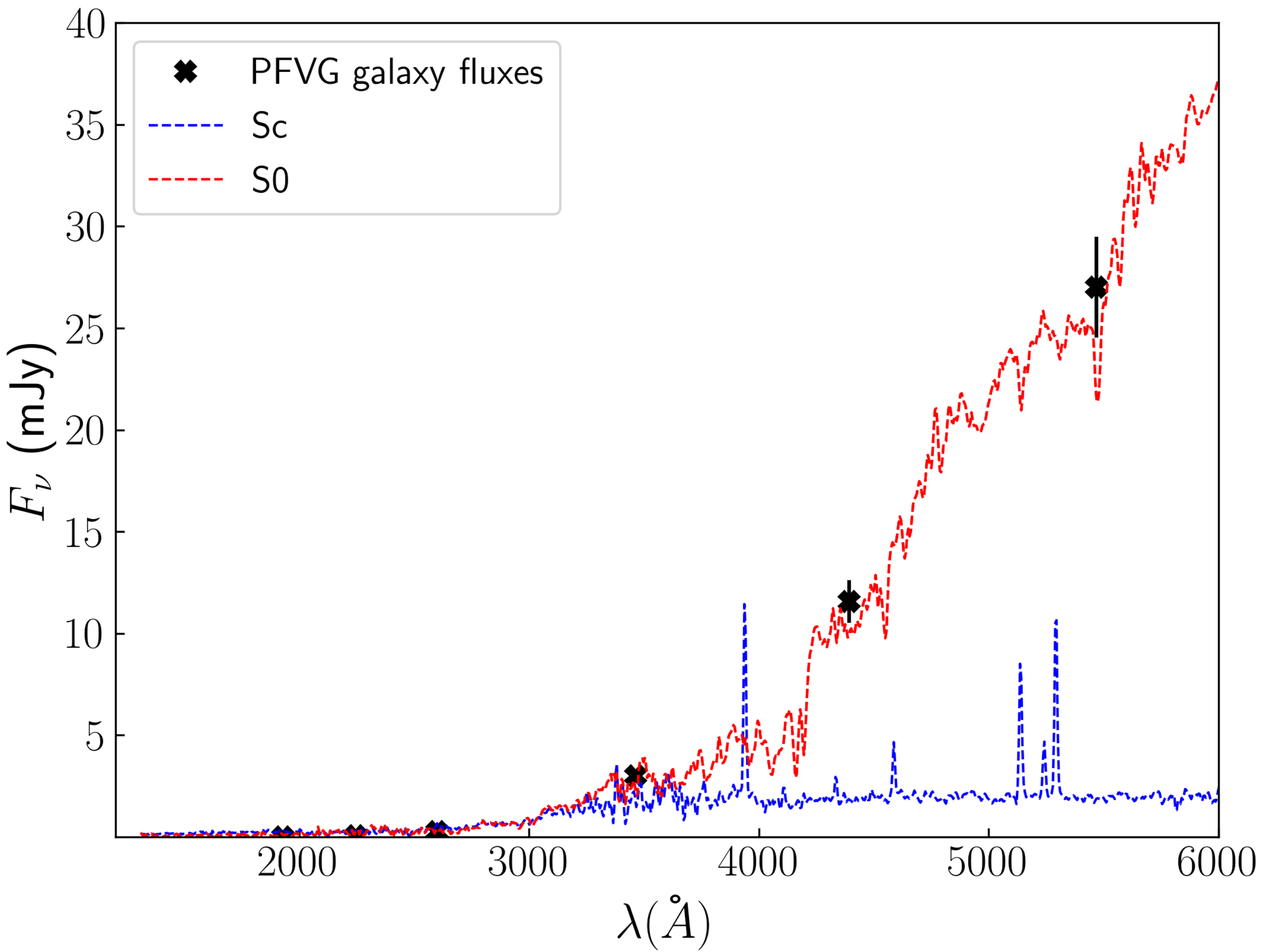}
\caption{Recovered PFVG host galaxy spectrum. The black stars correspond to the median of the PFVG distributions (Figure \ref{fvghost}) with error bars representing the $1\sigma$ uncertainty around the median. The PFVG method results in a host galaxy spectrum, which is consistent with an S0 galaxy template (red dotted line). For comparison, we show the template of an Sc galaxy morphology (blue dotted line). Both templates are from \citet{Kinney96}.}
\label{hostsed}
\end{figure}

Since the light curves are noisy, we estimated the density of possible intersections between the AGN and galaxy slope using Bayesian principal component analysis, which is included in the recently introduced probabilistic version of the FVG (PFVG\footnote{https://github.com/
HITS-AIN /ProbabilisticFluxVariationGradient.jl/}; \citealt{2022A&A...657A.126G}).
Figure \ref{fvghost} shows the recovered distribution of the host galaxy flux for each photometric band. As expected, the contribution of the galaxy to the total flux is negligible at the shorter wavelengths (< 4\%). However, at the longer wavelengths (U, B and V), the contribution is significant, ranging from 18 to 70\% of the total flux. Figure \ref{hostsed} shows the recovered spectral energy distribution (SED) of the host galaxy. The SED is consistent with the morphology of an S0 host galaxy. The host galaxy and AGN fluxes are given in Table \ref{fvgsumm}.

To estimate \iras\ bolometric luminosity, we applied the $5100$\AA\ bolometric correction of \cite{2017MNRAS.470..800T} as presented in \cite{2020A&A...635A..92G},

\begin{equation}
\log\,\left( \frac{L_{\mbox{\scriptsize Bol, 5100 \AA}}}{\ergs} \right) 
= 0.916\, \log\,\left( \frac{\nu L_\nu(\mbox{5100 \AA})}{\ergs} \right) 
+ 4.596.
\label{bolumco}
\end{equation}
where $\nu L_\nu(\mbox{5100 \AA})$ is the monochromatic $5100$\AA\ AGN luminosity. We obtained $\nu L_\nu(\mbox{5100 \AA})$ from the interpolation of the host-subtracted AGN fluxes assuming a power law $F_{\nu} \propto \nu^{1/3}$ SED. At a distance of 257.33 Mpc\footnote{We assumed a concordance cosmology with ${H_{0}=70\ \mathrm{km\ s^{-1}\ Mpc^{-1}}}$, $\Omega_{\Lambda}=0.73$ and $\Omega_{m}=0.27$.}, this gives an AGN luminosity of $\nu L_\nu(\mbox{5100 \AA}) = (3.41 \pm 1.20)\times 10^{44}{\mathrm{erg\ s^{-1}}}$. Using equation~\ref{bolumco} yields a bolometric luminosity $L_{\rm{Bol}} = 2.44\times 10^{45}{\mathrm{erg\ s^{-1}}}$. Assuming a  mass to radiation conversion efficiency $\eta = L_{\rm Bol}/\dot M c^2 = 0.10$ (\citealt{2009ApJ...690...20S}), we estimate the mass accretion rate $\dot M = 0.43 M_{\odot} yr^{-1}$. This result is consistent with $\dot M = 0.44 M_{\odot} yr^{-1}$ obtained by \cite{GRAVITY20}.% given the large uncertainty (+0.3/-0.4 dex) in the reported black hole mass.

\begingroup
\setlength{\tabcolsep}{4.5pt} % Default value: 6pt
\renewcommand{\arraystretch}{1.5} % Default value: 1
\begin{table}
\begin{center}
\caption{Foreground extinction, host galaxy and AGN fluxes.}
\label{fvgsumm}
\begin{tabular}{@{}cccc}
\hline\hline
Band & Extinction & Galaxy & AGN \\
             &  (mag) & (mJy) &  (mJy) \\   
\hline
UVW2 &  1.04 & $0.03\pm0.002$ & $7.78\pm0.18$ \\
UVM2 &  0.99 & $0.15\pm0.01$    & $9.21\pm0.27$ \\
UVW1 &  0.93 & $0.56\pm0.05$    & $12.99\pm0.31$ \\
U          &  0.79 & $4.20\pm0.39$  & $18.77\pm0.70$\\
B          &  0.65 & $12.85\pm1.18$  & $14.79\pm1.38$\\
V          &   0.49 &  $25.79\pm2.37$ & $9.53\pm2.63$\\
\hline
\end{tabular}
\end{center}
\end{table}
\endgroup

\begin{figure*}
 	\includegraphics[width=19cm, clip]{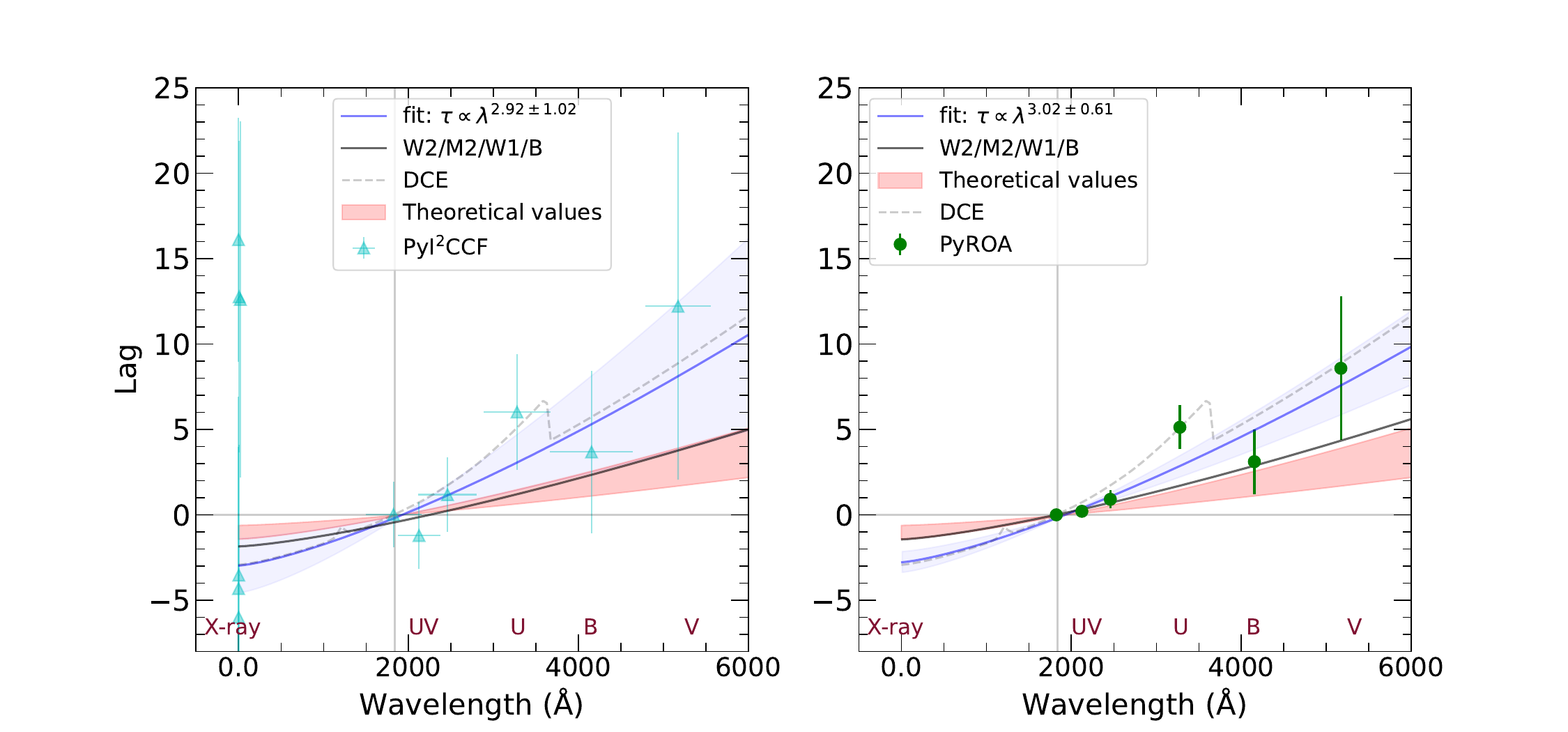}
\caption{Time delay spectrum determined with PyI$^2$CCF (left) and PyROA (right). The blue line corresponds to the best fit found by the relation $\tau \propto \lambda^{4/3}$. The red shaded area shows the predicted theoretical estimates for the size of the accretion disc. The black solid line is the fit \textbf{excluding} the contribution from the DCE.
}
    \label{fig:fit}
\end{figure*}

\subsection{Accretion disc modelling}
\label{ADmodeling}

The time delay spectral analysis was carried out considering the standard AGN accretion disc theory. The temperature profile of the accretion disc follows the relation $T(R) \propto R^{-3/4}$, and with the wavelength varying with temperature by the form $\lambda \propto T^{-1}$ and the delay $\tau = R /c$, we can derive a relationship between the delay and the wavelength of the form, 

\begin{equation}
\label{eqn:cont_lag}
\tau = \tau_{0}\left(\frac{\lambda}{\lambda_{0}}\right)^{\beta} - y_{0}
\end{equation}
where $\lambda_{0}$ is the reference wavelength corresponding to the \w-band ($\lambda 1928$ \AA), $\tau_{0}$ is the reference time delay which measures the radius of the disc emitting at the reference wavelength $\lambda_{0}$, and $\beta$ is the characteristic exponential of equation~\ref{eqn:cont_lag} which reflects the disc temperature profile. Figure~\ref{fig:fit} depicts the time delay spectrum obtained with PyI$^2$CCF and PyROA in the rest-frame. We fit the data using equation~\ref{eqn:cont_lag}, following the methodology presented by \cite{GonzalezB2022}. Five fits were made for the time delays obtained with PyI$^2$CCF. In the first two cases, the UVOT bands plus the hard X-ray delay (HX + UVOT) were considered because their value is negative and  follows the lamp post model.  {\sc javelin} data were not fitted, because the error bars are small and perhaps overestimated, and their results are unreliable.

\begin{table*}
	\centering
	\caption{Parameters estimated from the fits to the time delay spectrum. Column 3: shows the bands used for the fit, all-data corresponds to the X-ray data up to the V-band, and UVOT corresponds to the \w\, to V bands. Columns 4,5,6 correspond to the fit parameters obtained from equation~\ref{eqn:cont_lag}. Column 7: $\chi^2$/degrees of freedom.}
	\label{tab:lagfitinfo}
	\begin{tabular}{lllcccc} % four columns, alignment for each
		\hline\hline
		Fit & Code  &Data & $\tau_0$ & $\beta$ & $y_{0}$ & $\chi_\nu^2$\\ 
               &   &   & (days) &    \\  
         (1)  &  (2)   & (3) & (4) & (5) & (6) & (7) \\      
         \hline
         & {\tt PyI$^2$CCF}& & &  &  & \\
		(a) & &X-ray + UVOT & 2.97$\pm$0.91 & 4/3 & 1.00 & 0.33 \\
		(b) & &UVOT &  2.92$\pm$1.02 &  4/3 & 1.00 & 0.40\\
		(c) & &X-ray + UVOT & 3.27$\pm$1.02 & 4/3 & 1.12$\pm$0.18 & 0.31 \\	%		\hline
		(d) & &UVOT &  2.97$\pm$1.16 &  4/3 & 1.12$\pm$0.24 & 0.44\\
		(e) & &W2/M2/W1/B & 1.50$\pm$1.32 & 4/3 & 1.22$\pm$0.36 & 0.24 \\	\hline
		& {\sc PyROA}& & &  &  & \\
		(f) &  & UVOT & 3.02$\pm$0.61 & 4/3 & 1.00 & 1.95\\
		(g) & & UVOT & 2.70$\pm$0.66 & 4/3 & 0.92$\pm$0.00 & 16.33 \\
		(h) & &W2/M2/W1/B & 1.47$\pm$1.12 & 4/3 & 1.23$\pm$0.37 & 0.24 \\
   \hline\hline 
	\end{tabular}
\end{table*}

\begin{table*}
	\centering
	\caption{Fit parameters for FFPs.}
	\label{tab:fitxray}
	\begin{tabular}{clccc|ccc} % four columns, alignment for each
		\hline\hline
		Light Curve&Band & $\alpha$ & $\beta$ & $\chi_v^2$ & $m$ & $b$ & $\chi_v^2$   \\
		 &    &   &   &  \\
		 Number & $keV$     &  &  &  \\
        (1)     & (2) & (3) & (4) & (5) & (6) & (7) & (8) \\
		\hline
        \hline
        1 & 0.3 - 1.0  & 0.07$\pm$0.01 & 0.27$\pm$0.06 & 1.16 & 0.15$\pm$0.01 & 0.02$\pm$0.06 & 1.26 \\
        2 & 1.0 - 2.0  & 1.00$\pm$0.00 & 1.00$\pm$0.00 & 0.00 & 1.00$\pm$0.00 & 0.00$\pm$0.00 & 0.00 \\
        3 & 2.0 - 3.0  & 0.30$\pm$0.06 & 0.65$\pm$0.07 & 2.91 & 0.54$\pm$0.06 & 0.01$\pm$0.07 & 3.01 \\
        4 &3.0 - 4.0  & 0.12$\pm$0.03 & 0.38$\pm$0.08 & 4.53 & 0.27$\pm$0.03 & 0.02$\pm$0.08 & 4.83 \\
        5 &4.0 - 5.0  & 0.06$\pm$0.01 & 0.27$\pm$0.07 & 2.23 & 0.14$\pm$0.01 & 0.02$\pm$0.07 & 2.37 \\
        6 &5.0 - 6.0  & 0.03$\pm$0.01 & 0.25$\pm$0.10 & 1.94 & 0.07$\pm$0.01 & 0.01$\pm$0.01 & 1.96 \\
        7 & 6.0 - 7.0  & 0.01$\pm$0.004 & 0.12$\pm$0.10 & 1.19 & 0.01$\pm$0.004 & 0.007$\pm$0.10 & 1.20 \\
        8 & 7.0 - 8.0  & 0.01$\pm$0.008 & 0.40$\pm$0.17 & 1.58 & 0.03$\pm$0.008 & 0.002$\pm$0.17 & 1.57 \\
        9 & 8.0 - 9.0  & 0.002$\pm$0.001 & -0.02$\pm$0.15 & 140.6 & -0.003$\pm$0.001 & 0.00$2\pm$0.15 & 1.42\\
        10 & 9.0 - 10.0  & 0.005$\pm$0.003 & 0.36$\pm$0.21 & 0.19 & 0.01$\pm$0.003 & 0.001$\pm$0.21 & 0.19 \\
        \hline
	\end{tabular}
\end{table*}

We divided the analysis in two cases. In the first case, $\tau_0$ and $y_0$ were assumed to be free, while $\beta$ was fixed at 4/3, as predicted by standard AGN accretion disc theory. In the second test we only let $\tau_0$ vary, leaving the other two parameters fixed. Similar results were found in both fits. If we set $y_0$ to 1 (cases a and b), the resulting $\tau_0$ values for the HX-ray+UVOT data are 2.97$\pm$0.91 and 2.92$\pm$1.02, while the $\tau_0$ values for the UVOT data alone are 3.27$\pm$1.02 (case c) with a corresponding $\chi_v^2$ of 0.31 and 2.97$\pm$1.16 (case d) with a $\chi_v^2$ of 0.44. These fits showed good agreement. For the PyROA analysis, we fitted only the UVOT data, resulting in $\tau_0$ values of 3.02$\pm$0.61 (case f) and 2.70$\pm$0.66 (case g) when $y_0$ was set to 1.0 and 0.92$\pm$0.00, respectively. The corresponding $\chi_v^2$ values were 1.95 and 16.33. In cases (e) and (h), we considered only the \w/\m/\wb\, and $B$ points and obtained $\tau_0$ values of $1.50\pm1.32$ (with a $\chi_v^2$ of 0.24) for the PyI$^2$CCF data and $1.55\pm0.28$ (with a $\chi_v^2$ of 0.22) for the PyROA data. In all cases when $y_0$ was set to 1, the time delay of the reference light curve (UVW2) was not included, so the model did not deviate from 0. The results of the fitting are shown in Figure~\ref{fig:fit} and the best fit parameters are summarised in the table~\ref{tab:lagfitinfo}.

It is clear that the simple power law model given by equation \ref{eqn:cont_lag} cannot well reproduce the observed time delay. It particularly fails to mimic the strong jump seen at the $U$ band. As we mentioned in section \ref{lcandvariable}, this feature is most likely the result of contamination by diffuse continuum emission (DCE) from the BLR. DCE was theoretically predicted by \cite{Korista2001} and only recently its contribution has been observed thanks to RM campaigns with high temporal resolution targeting the accretion disc (\citealt{2018ApJ...857...53C,2020ApJ...896....1C}; \citealt{Edelson19}; \citealt{hernandez2020}).

\cite{Korista19} quantified the DCE contribution in the well-known AGN NGC5548 and provided a recipe to constrain the DCE in the delay spectrum of other AGN (see also \citealt{hernandez2020}; \citealt{2023MNRAS.tmp..408P}). To account for the difference in luminosity between NGC5548 and \iras, we scaled the DCE fractional contribution and the delay spectrum of NGC5548 using the BLR luminosity relation $R_{\rm BLR}\propto L_{\rm{ AD }}^{\alpha}$ ($\alpha = 0.533^{+0.035}_{-0.033}$; \citealt{bentz13}). The scaled DCE contribution was then added to the power law given by the equation \ref{eqn:cont_lag}. The new fit is shown in Figure~\ref{fig:fit} (black line). The addition of the DCE can clearly mimic the increase in delay towards the $U$ band, which is about 1.5 times larger than the simple power law.

In the next we compare the standard AGN accretion disc theory and the fits made to the time delay-wavelength relationship.
We estimate the average accretion disc size following \cite{Fausnaugh16} and \cite{Edelson19},

\begin{equation}
    r = 0.09\left(X\frac{\lambda}{\lambda_{0}} \right)^{4/3} M_{8}^{2/3} \left( \frac{\dot{m}_{Edd}}{0.10}\right)^{1/3} lt-days\,,
    \label{eq:sizeDisk}
\end{equation}
Where $X$ is a multiplicative factor of 4.97 if Wien's law gives the temperature $(T)$ of the wavelength $(\lambda)$ at a radius $r$, or equal to $2.49$ when the flux-weighted estimate assumes that the temperature profile of the disc is described by $T \propto R^{4/3}$. In our case, $\lambda_0$ corresponds to the driver wavelength, which is equivalent to the light curve of the \w-band with a wavelength of 1928\AA. $M_8$ corresponds to the black hole mass in units of $10^8\,\,M_\odot$, and $\dot{m}_{Edd}$ is the Eddington ratio ($L_{\rm bol}/L_{\rm Edd}$).  According to \cite{GRAVITY20}, the mass of the black hole is $1\times10^8 M_{\sun}$ and using the bolometric luminosity determined in section \ref{sec:FVG}, the theoretical prediction of the size of the accretion disc is 3 to 5 times smaller than the observations.

\begin{figure*}
 	\includegraphics[trim=0.2cm 10.0cm 1.0cm 10.0cm, clip, width=12cm]{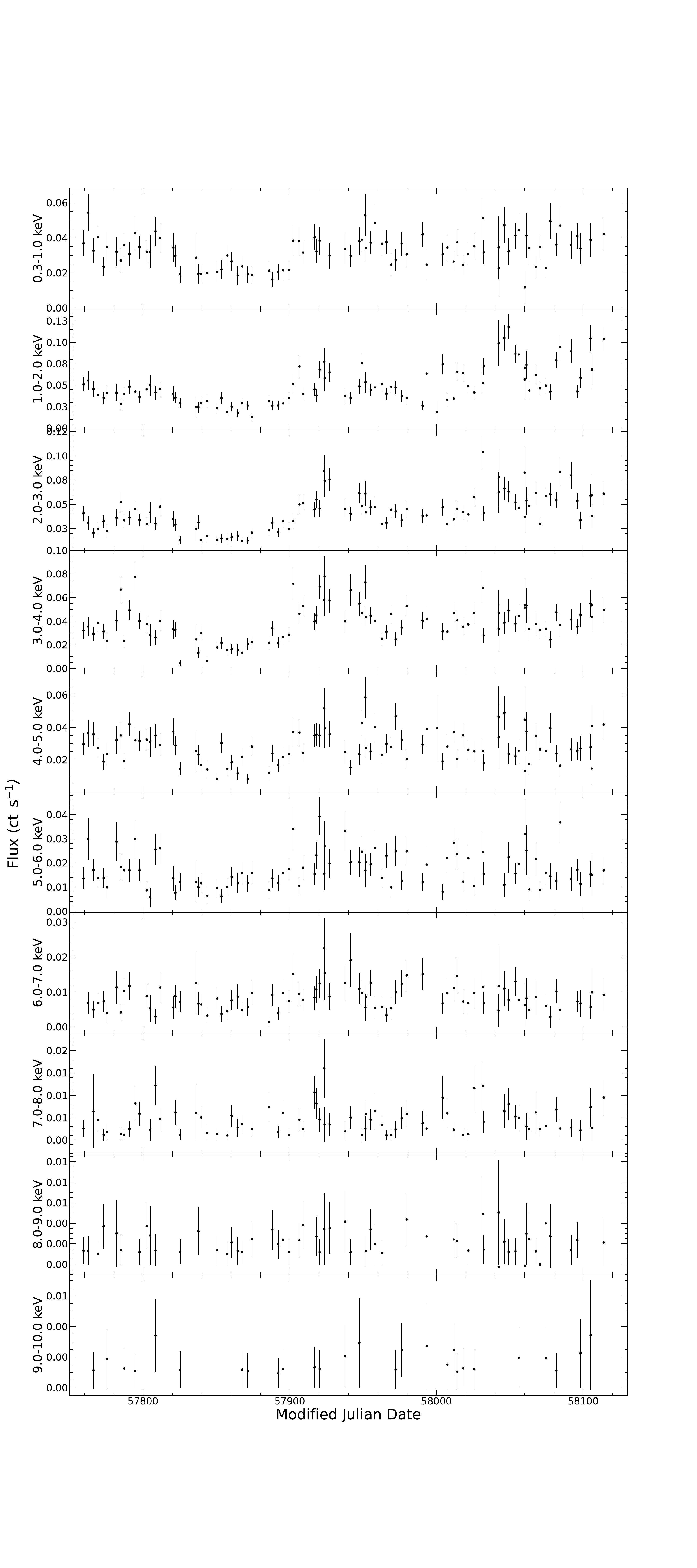}
\caption{X-ray light curves extracted in different energy ranges, with a delta width of 1 keV per light curve. The light curves are ordered from lowest energy (upper panel) to highest energy (bottom panel).}
    \label{fig:lc_xray}
\end{figure*}

\subsection{X-ray variability correlation}
\label{xrayvariability}

To investigate the temporal relationships between the X-ray light curves, we performed a simple correlation analysis using the cross-correlation coefficient (\citealt{Derrick2004}) without any interpolation of the light curves.
By performing a curve-by-curve cross-correlation analysis, we attempt to identify temporal relationships between the light curves and reveal possible patterns of variability.
For this purpose, we have extracted a total of 10 X-ray light curves from observations with Swift, covering an energy range from 0.3 to 10 keV (see section~\ref{sec:obs}). 
Each light curve corresponds to a specific energy range, with a width of about 1 keV. 
The first light curve covers the energy range of 0.3-1.0 keV, followed by subsequent ranges of 1.0-2.0 keV and so on, up to the 10th keV. 
In Figure~\ref{fig:lc_xray} we show the X-ray light curves arranged in different panels according to their energy ranges. 
The upper panel represents the light curve with the lowest energy range, while the lower panel corresponds to the light curve with the highest energy range.

\begin{figure}
  \centering
  \includegraphics[width=\columnwidth]{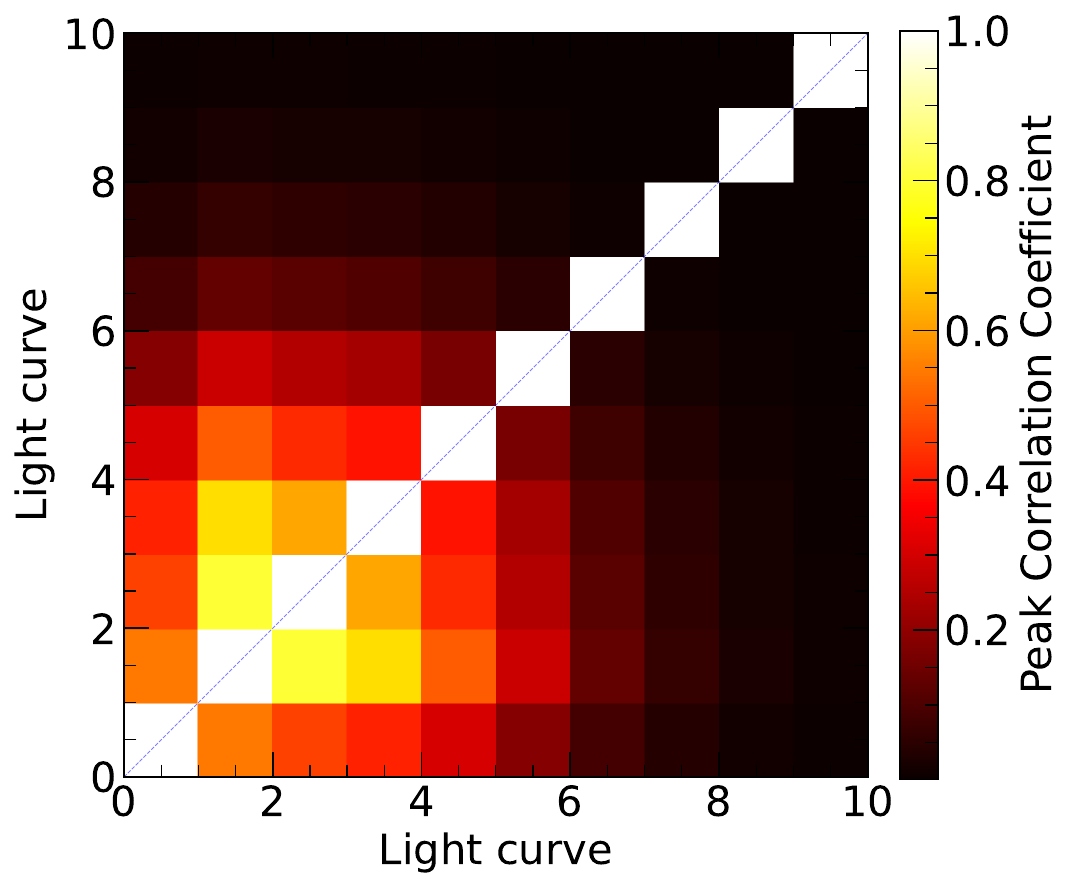}
\caption{Cross-correlation matrix of X-ray light curves. The matrix represents the cross-correlation values between the 10 X-ray light curves, and each number represents a light curve going from low to high energy as listed in column one of Table~\ref{tab:fitxray}. Intense colours (e.g. red, yellow and white) indicate a higher correlation, while opaque colours (e.g. black) indicate a lower correlation. The matrix shows the strongest correlations between X-ray light curves with lower energy ranges, which gradually decrease as the energy range increases.}
\label{fig:crossXray}
\end{figure}

The light curves were read one by one, and a cross-correlation matrix was created to store the cross-correlation values between pairs of light curves.
This matrix has dimensions of $N\times N$, where $N$ represents the number of X-ray light curves.
The dimensions of the matrix allow a comprehensive analysis of the correlations between all possible pairs of light curves.
Figure~\ref{fig:crossXray} shows the resulting cross-correlation matrix for the 10 X-ray light curves, where the light curves in the matrix are represented by the numbers given in Table~\ref{tab:fitxray}. The higher correlation is represented by more intense colours such as red, yellow and white. Conversely, light curves with lower correlation are represented by opaque colours such as black.
It is evident that the correlation between X-ray light curves decreases with increasing energy range.
Notably, the correlation coefficient for the light curve in the 0.3-1.0 keV energy band is close to 1 and for the 2.0-3.0 keV light curve (shown in white in Figure~\ref{fig:crossXray}) is 1 as it correlates with itself.
As we move to higher energy ranges, the correlation coefficient gradually decreases, reaching about 0.2 for the light curve of 7.0-8.0 keV.
This indicates a strong similarity of variability behaviour between these light curves.
However, for the light curves in the 8.0-10 keV range, the correlation coefficient is practically zero.
Thus, we observe a correlation between the light curves up to the energy range of 8.0 keV, indicating that the variabilities observed in the energy ranges from 0.3 to 8 keV are characteristic of AGN.
In contrast, no correlation is observed for the hard X-ray light curves in the energy range from 8 to 10 keV, and no clear AGN-related variability is evident.
The strong correlation within certain energy bands suggests consistent patterns of variability related to specific physical processes in the AGN.
The gradual decrease in correlation with increasing energy suggests a shift in the dominant mechanisms driving X-ray variability.

\begin{figure*}
 	\includegraphics[trim=0.2cm 7.8cm 1.0cm 8.0cm, clip, width=15cm]{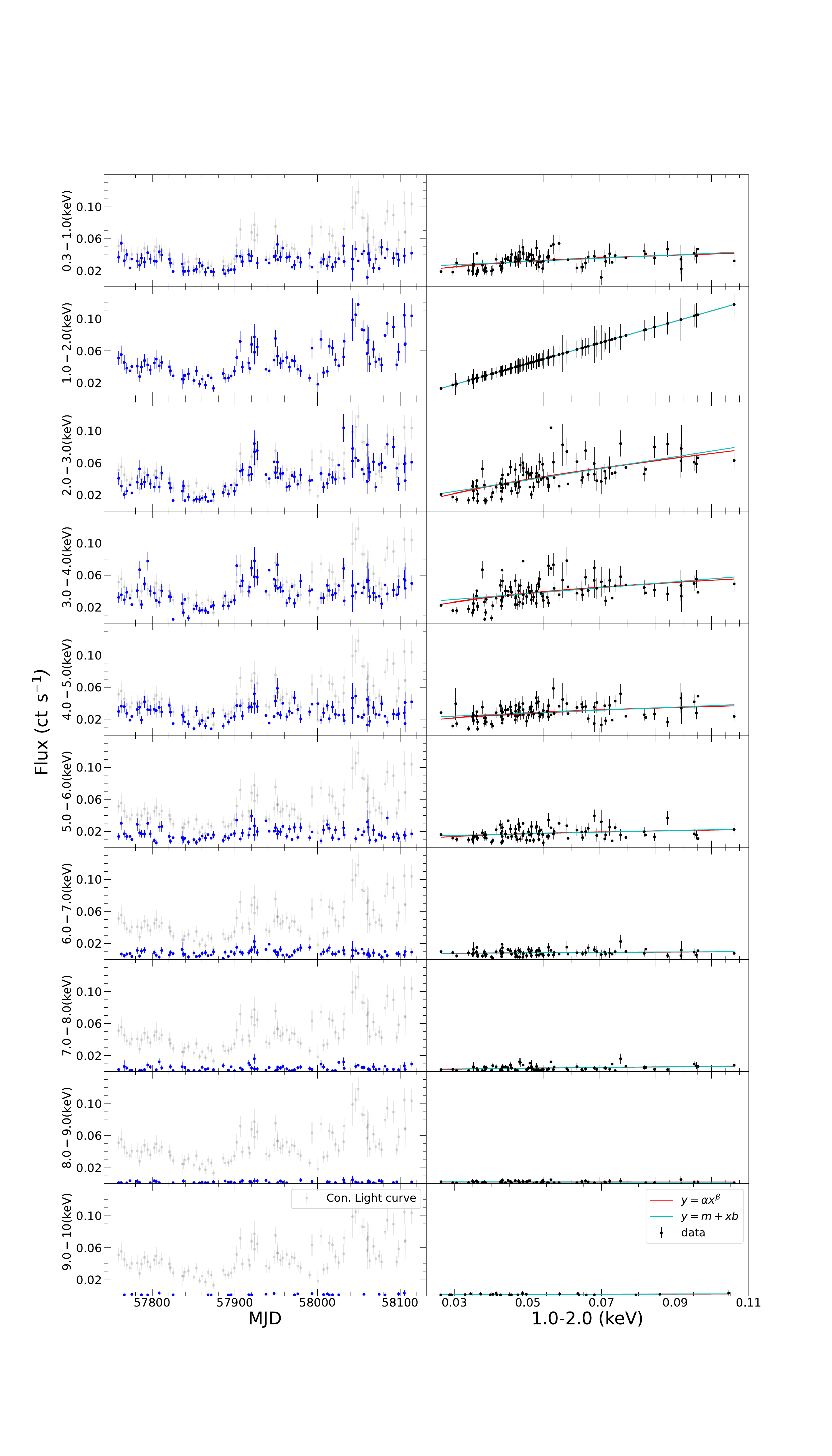}
\caption{Left panel: X-ray light curves for different energy ranges. The blue points with error bars correspond to the data of the light curves extracted for energy ranges equivalent to 1 keV. The faded black points corresponds to the  light curve obtained in the energy range of 1.0-2.0 keV. Right panel: Results of the FFP analysis for different X-ray energy ranges. The red line corresponds to the best power law fit, and the cyan line indicates the best fit obtained with a linear model.}
    \label{fig:FFP}
\end{figure*}

\subsection{Flux-Flux Plot analysis}
\label{ffpanalysis}

In order to examine the behaviour of the X-ray variability, we have used the Flux-Flux Plot method (FFP) (e.g., \citealt{2011PASJ...63S.925N,2013PASJ...65....4N}; \citealt{Kammoun15}). This method is used to distinguish modes of spectral variability, and it is characterised by a linear model $y=bx+m$ and a power-law $y=\alpha X^{\beta}$.

To generate the FFP, we employed the 10 extracted X-ray light curves (see Figure~\ref{fig:lc_xray}). For this analysis, we selected the light curve in the 1-2 keV energy range as the reference curve due to its prominent and well-defined variability characteristics. We contrast it with the other light curves in Figure~\ref{fig:FFP}. We then perform two fits to the FFP, a linear and a power fit, where the parameters $\alpha$, $\beta$, $m$ and $b$ are free to vary. $X$ and $Y$ represent the count rate in the light curves of the different extracted bands. Both fits are shown in Figure~\ref{fig:FFP}. Table~\ref{tab:fitxray} shows the results of $\alpha$, $\beta$, $m$ and $b$ along with the $\chi^2$ for the fit of each FFP.

\begin{figure}
  \centering
  \includegraphics[width=\columnwidth]{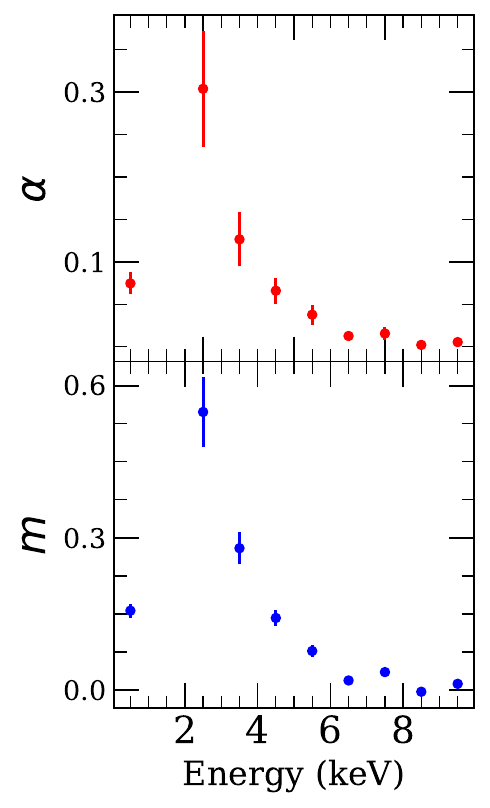}
\caption{Reconstructed time-averaged spectrum for the non-variable component using the power law (top) and the linear model (bottom).}
\label{fig:FFPresult}
\end{figure}

The FFP analysis shows a weak positive correlation between the reference light curve at 1.0-2.0 keV and the softer energy range at 0.3-1.0 keV, with a Pearson correlation coefficient of $P = 0.4$.
The correlation is strongest in the energy range of 2.0-3.0 keV, with $P = 0.6$.
However, it begins to weaken at about 3.0-4.0 keV ($P = 0.4$) and continues to weaken with increasing energy until there is no correlation at all in the hardest energy range of 9.0-10 keV ($P = 0$).
In the low energy regions, between 0.3 and 6.0 keV, both the constant part of the linear model ($m$) and the power law ($\alpha$) are non-zero and positive (see table~\ref{tab:fitxray}). This indicates the presence of a non-variable spectral component, which in this case ranges from 0.3 to 6.0 keV. This softer constant spectral component has been observed before in other AGN (e.g., \citealt{2011PASJ...63S.925N}). The linear and power law models give a very similar shape for the constant spectral component, as shown in Figure \ref{fig:FFPresult}.

Another interesting feature is the presence of a bulge or excess emission in the same lower energy ranges of 0.3 to 6.0 keV as described above.
The $\chi^2$ for these fits are high due to the dispersion of points (see table~\ref{tab:fitxray}), showing that the Flux-Flux relationship is not linear and cannot be described by a simple power law model, assuming a more intrinsic variability in which the multiple components can vary simultaneously.
Despite the fact that the excess emission is observed across the entire energy range 0.3-6.0 keV, the highest excess is observed at the energy level where the correlation is strongest ($P = 0.6$), implying that the excess is produced in a region which is well constrained by the energy ranges 2.0-3.0 keV. Using high-resolution XMM-Newton and NuSTAR observations a year later than {\em Swift}, \cite{Walton2020} discovered a strong relativistic reflection of the iron K$\alpha$ line at $\sim 6$ keV in the spectrum of \iras. The relativistic iron line has a long tail extending up to about 3.0 keV. It could be that the variability due to excess emission that we observe in the FFP analysis is partly due to this relativistic reflection in the inner part of the accretion disc. The relativistic K$\alpha$ line peaks at 6.4 keV and at this energy level our FFP shows the excess, but with a much lower amplitude (see Figure~\ref{fig:FFP} for 5.0-6.0 keV) compared to the 2.0-3.0, 3.0-4.0 and 4.0-5.0 keV regions. The relativistic K$\alpha$ line dominates between 3.5 and up to about 6.8 keV, but not between 2.0 and 3.0 keV. Therefore, it can only partially contribute to the total excess seen in the FFP in the softer energy range.

According to the standard lamp-post model, such reflection should lead the UV/optical emission.
However, we observe that the UV leads the soft X-ray emission by about 15 days (see Figure \ref{fig:lag_data}).
From the light curves (Figure \ref{fig:LagJavelin}) we see that the flux variations of the soft X-rays (SX) is well correlated with the UV (UVW2) variations until almost the end of the campaign. It is only from MJD58000 onwards that interesting features start to appear. We see, for example, that the soft X-ray flux increases sharply from MJD58000 until it reaches a maximum at MJD58040.
This maximum and the beginning of the increase of the flux in the UV are clearly behind the flux of soft X-rays by about 20 days, from MJD57980 to MJD58020.
This is clearly the strongest delayed feature detected in the cross-correlation analysis.
After the peak observed in the soft X-rays at MJD58040, the flux decreases abruptly until it increases again at about MJD58060, reaching its final maximum at MJD58100. This is again about 20 days later than the secondary maximum observed in the UV at MJD58080. JAVELIN does not provide a good fit for the second maximum at the end of the campaign, but the traditional cross-correlation function shows a clear peak at about 20 days.

A similar behaviour of soft X-rays was recently reported by \cite{2023arXiv230207342K} for the source Mrk335. The authors found a sharp increase in UV flux preceding soft X-rays by about 15 days. They refer to this increase in flux as a flare, which was also seen in the hard X-ray band. Two different scenarios have been proposed to explain the long delay: a) the occurrence of a later inflow in the disc, caused by a pressure or magnetic wave that triggers the sharp increase in UV flux, followed by the fluctuations of the soft X-rays. In this case, the 15-day time scale is consistent with dynamical rather than viscous time scales, and b) a late expansion of the corona together with a partial obscuration of the line of sight, so that the observer sees the early UV response of the disc followed by the soft X-ray emission (see Figure 8 in \citealt{2023arXiv230207342K}). Such a scenario would be more consistent with what we see in \iras . The excess emission could be the result of a more extended corona expanding along the black hole's rotation axis, rather than a point source. The expansion of the corona could be triggered by instabilities in the inner region of the accretion disc that occurred at a later time. If the corona expands, it would heat up the disc even more. This scenario of an expanded corona was recently studied in detail by \cite{2023MNRAS.520..180H}. More distant material above the disc could block the observer's line of sight, so that the observer first sees the increase in the disc's UV/optical emission, which is less affected by scattering than the soft X-ray emission. This scenario would require an unknown scattering source above the disc, at a greater distance from the inner region. Whether this scattering source is an obscurer, as observed in Mrk335, cannot be answered with our current data. We dedicate the modelling and more detailed analysis of this long time delay to a later article, after we have collected additional data with higher spectral resolution.

\section{conclusions}

We have carried out a reverberation mapping study of the AGN \iras\ using multi-wavelength observations, including X-rays, UV and optical performed with the {\em Swift} telescope. The main results are:

\begin{enumerate}

\item Time delays determined with three different algorithms PyI$^2$CCF, PyROA, and JAVELIN provide consistent results. The UV/optical time delay 
increases with wavelength and ranges from values close to zero for the short wavelength bands (\m,\wb -bands) to about 11 days for the long wavelengths 
($V$-band). This shows a trend similar to that reported in previous reverberation mapping studies (e.g., \citealt{Fausnaugh16,Edelson19,pozo19,hernandez2020,GonzalezB2022}).

\item Flux variation gradient analysis suggests that the host galaxy may have an S0 morphology. The contribution of the galaxy is negligible at short wavelengths, while at long wavelengths it can reach up to about 70\% of the total flux. The derived bolometric luminosity of 3.41$\times10^{44}\,erg\,\,s^{-1}$ with a mass accretion rate of 0.43$M_{\odot}\,yr^{-1}$ is consistent with \cite{GRAVITY20}.

\item After corrections for the host galaxy, extinction and diffuse continuum emission of the BLR, the time-delay spectral analysis shows that the size of the accretion disc is up to five times larger than that predicted by standard AGN accretion disc theory.

\item The X-ray bands exhibit a special behaviour. The calculated time delay for the hard X-ray band is negative. In contrast to the lamp-post model prediction, the soft X-ray light curve lags behind the UV, with a delay of $13\pm4$ days obtained with PyI$^2$CCF and $5\pm0.3$ days obtained with JAVELIN. The analysis of the X-ray variability with the FFP method shows a contribution of a non-variable spectral component in the ranges from 0.3 to 6.0 keV. The shape of the constant spectral component is similar to other AGN. However, an additional bulge or excess emission is detected, which could be partly attributed to relativistic reflection in the inner part of the accretion disc. According to the lamp-post model, this reflection should lead the UV variability. Perhaps \iras\ is another example of a source with an expanding corona rather than a point source. However, more detailed observations are needed to confirm this scenario.

\end{enumerate}

\section*{Acknowledgements}

DHGB acknowledges CONACYT support $\#$319800 and of the researchers program for Mexico. 
FPN gratefully acknowledge the generous and invaluable support of the Klaus Tschira Foundation. 
FPN acknowledges funding from the European Research Council (ERC) under the European Union's Horizon 2020 research and innovation programme (grant agreement No 951549).
This research has made use of the NASA/IPAC Extragalactic Database (NED) which is operated by the Jet Propulsion Laboratory, California Institute of Technology, under contract with the National Aeronautics and Space Administration. This research has made use of the SIMBAD database, operated at CDS, Strasbourg, France. We thank the anonymous referee for the constructive comments, which have greatly improved this paper. In honor of Pompilio Gonzalez 1941 - 2022.

%%%%%%%%%%%%%%%%%%%%%%%%%%%%%%%%%%%%%%%%%%%%%%%%%%
\section*{Data Availability}

The data used in this work is publicly available via the {\em Swift} Telescope Database.

%%%%%%%%%%%%%%%%%%%% REFERENCES %%%%%%%%%%%%%%%%%%

% The best way to enter references is to use BibTeX:

\bibliographystyle{mnras}
\bibliography{bibliography} % if your bibtex file is called example.bib

%%%%%%%%%%%%%%%%%%%%%%%%%%%%%%%%%%%%%%%%%%%%%%%%%%

%%%%%%%%%%%%%%%%% APPENDICES %%%%%%%%%%%%%%%%%%%%%

\appendix

\section{light curve}

\begin{figure*}
 	\includegraphics[trim=0.2cm 10.0cm 1.0cm 10.0cm, clip, width=12cm]{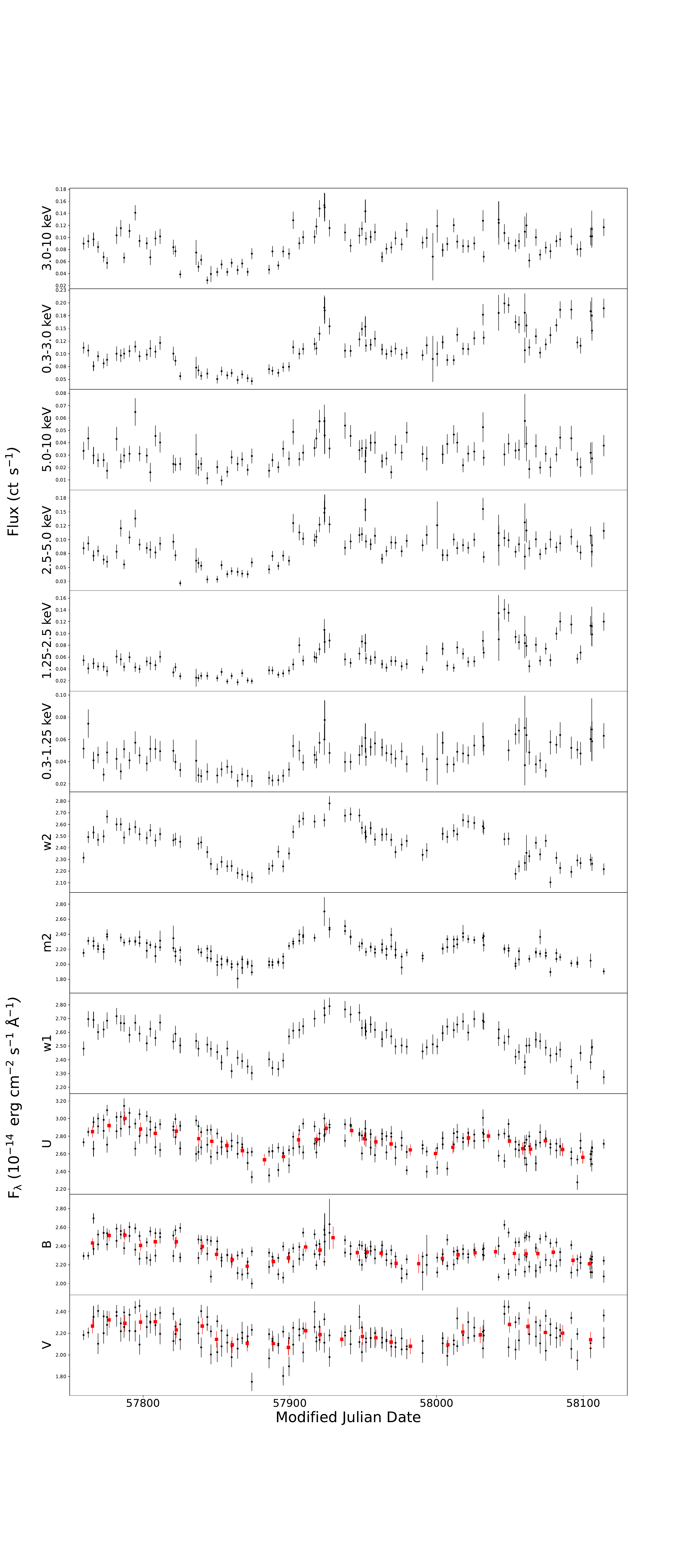}
\caption{X-ray, UV and optical light curves. Light curves are ordered from longest (upper panel) to shortest (lower panel) wavelength. the black points and errors correspond to all the data obtained in the observations. The red points and errors correspond to the average of 4 consecutive observation points. The upper panels correspond to X-rays light curves extracted in different energy ranges. The energy range is specified in the y-axis.
} \label{fig:LCall}
\end{figure*}

Figure~\ref{fig:LCall} shows the light curves in descending order from X-ray (upper panel) to optical (lower panel). The upper panels correspond to the X-ray light curves extracted in different energy ranges, corresponding to 3-10 keV which was considered as hard X-rays, and 0.3-3 keV considered to be soft X-rays. In addition, X-ray curves were extracted in smaller energy ranges, 5.0-10, 2.5-5-0, 1.25, 2.5, and 0.3-1.25 keV.

\begin{figure*}
 	\includegraphics[trim=0.2cm 0.0cm 0.0cm 0.0cm, clip, width=17cm]{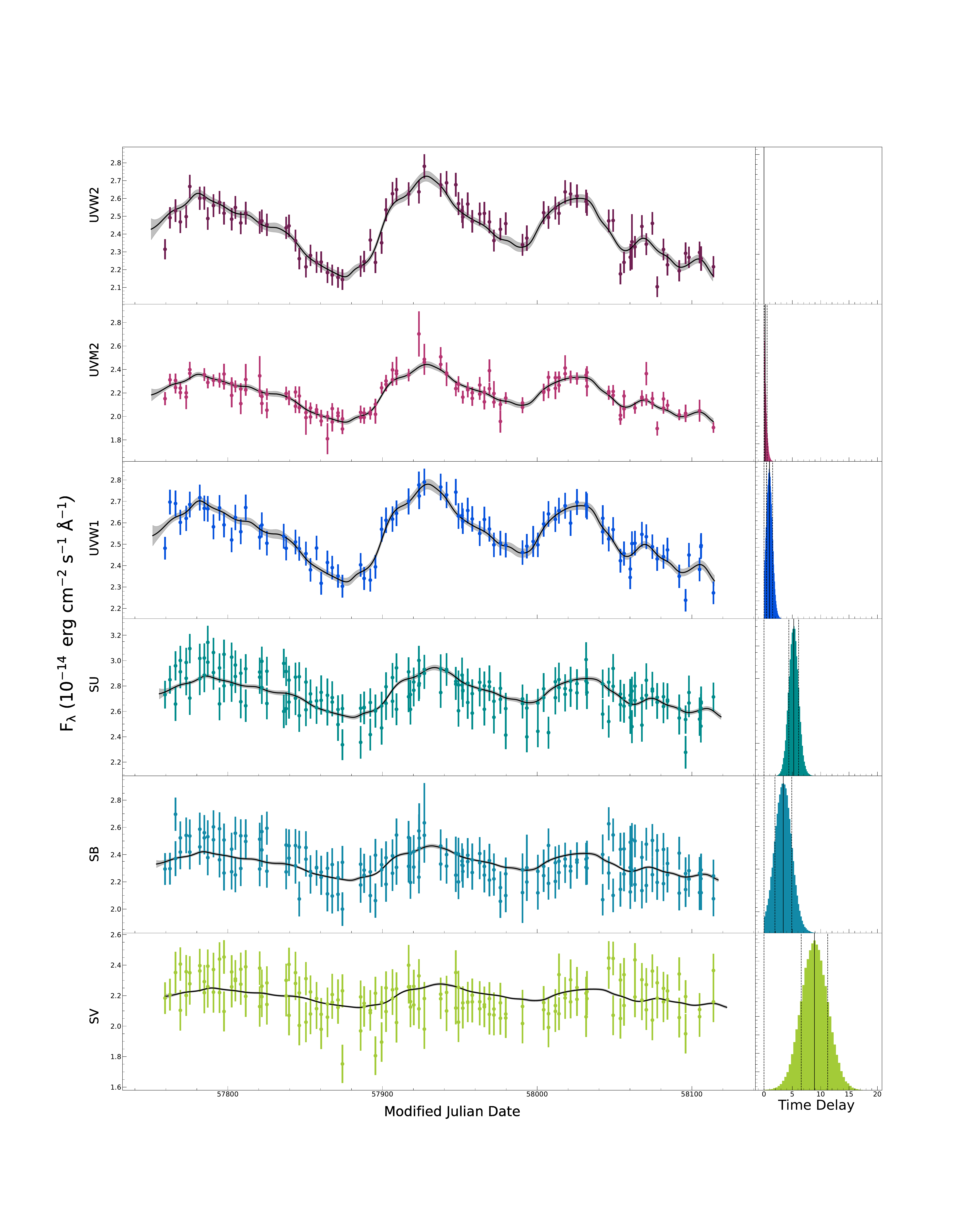}
\caption{Light curve models (left) and distribution of time delays (right) obtained with PyROA.}
    \label{fig:pyroa}
\end{figure*}

Figure~\ref{fig:pyroa} shows the \iras\, light curves obtained
with the six UVOT filters, together with the lag time measurements and fits made with the PyROA code.

%%%%%%%%%%%%%%%%%%%%%%%%%%%%%%%%%%%%%%%%%%%%%%%%%%

% Don't change these lines
\bsp	% typesetting comment
\label{lastpage}
\end{document}